\documentclass{cta-author}

\usepackage{xspace,amsmath,amssymb,epsfig,syntonly}
\usepackage{xcolor}

\newtheorem{theorem}{Theorem}{}
{}
\newtheorem{remark}{Remark}{}

\newtheorem{proposition}{\hspace{-11pt}\bf Proposition}

\begin{document}

\supertitle{Submitted to IET Communications}

\title{DEBIT: Distributed Energy Beamforming and Information Transfer for Multiway Relay Networks}

\author{\au{Zhaoxi Fang$^{1\corr}$},   \au{Xiaojun Yuan$^{2}$}}

\address{\add{1}{School of Electronics and Computer, Zhejiang Wanli University, Ningbo, China.}
\add{2}{ the National Key Laboratory of Science and Technology on Communications, University of Electronic Science and Technology of China, Chengdu, China.}
\email{zhaoxifang@gmail.com}
}

\begin{abstract}
In this paper, we propose a  new distributed energy beamforming and information transfer (DEBIT) scheme for realizing  simultaneous wireless information and power transfer (SWIPT) in  multiway relay networks (MWRNs), where multiple single-antenna users exchange information via an energy-constrained single-antenna relay node.   We investigate the optimal transceiver designs to maximize the achievable sum-rate or the harvested power. The resultant sum-rate maximization problem is non-convex and the global optimal solution can be obtained through a three-dimensional search in combination with conventional convex optimization. To reduce the computation complexity, a suboptimal DEBIT scheme is also proposed, for which the optimization problem becomes linear programming.  The achievable sum-rate performance is analyzed and a closed-form lower bound is derived for the   MWRN with a large number of users. Furthermore, we consider the harvested-power maximization problem under a target sum-rate constraint, and derive a lower bound of the average harvested power for MWRNs with a large number of users. Numerical results show that the  DEBIT scheme significantly outperforms the conventional SWIPT and the derived lower bounds are tight.
\end{abstract}

\maketitle

 %%%%%%%%%%%%%%%%%%%%%%%%%%%%%%%%%%%%%%%%%%%%%%%%%%%%%%%%%%%%%%%%%%%%%%
%                                                                    %
%               Section: Introduction                                %
%                                                                    %
%%%%%%%%%%%%%%%%%%%%%%%%%%%%%%%%%%%%%%%%%%%%%%%%%%%%%%%%%%%%%%%%%%%%%%

\section{Introduction}
In many   wireless networks, terminals are usually equipped with fixed power supplies and have a limited lifetime, e.g., sensor nodes embedded in buildings are equipped with batteries which are highly inconvenient to  replace. Harvesting energy from the environment has emerged as a promising solution for prolonging the lifetime of energy-constrained devices in wireless communication systems \cite{DYang15,Shin16,Varshney08,Grover10,Zhai16,Mu17}.  Considering the fact that radio frequency (RF)  signals carry information as well as energy at the same time, simultaneous wireless information and power transfer (SWIPT) technology has   gained considerable attention in both academic and industrial fields \cite{Zhou12,Zhang13,Chen16,Chen15,Wang13}.

The concept of SWIPT was first proposed in \cite{Varshney08} and \cite{Grover10}. In a SWIPT enabled wireless network, the receivers can process information and harvest energy from received signals simultaneously.   The SWIPT technique has been applied to various wireless networks, such as multi-input multi-output (MIMO) systems \cite{Zhang13,Wang15,WangF15}, orthogonal frequency division multiple access  (OFDMA) systems  \cite{Ng13},   wireless relay networks \cite{Liu16,Li13,Tutuncuoglu13}, and cognitive radio networks \cite{Lee13}. In \cite{Ng13}, the authors investigated the optimal resource allocation to maximize the energy efficiency for SWIPT OFDMA systems with power-splitting receivers. In \cite{Liu16}, the authors  analyzed the performance of SWIPT enabled two-way relay network  (TWRN), where the relay   harvests power from the sources' transmit signal based on a time-switching protocol.  In \cite{Tutuncuoglu13}, the authors investigated the optimal transmission policy for the SWIPT TWRN where both the users and the relay node are able to process information and harvest power simultaneously.

A key observation is that, in wireless communications, the electromagnetic wave impinged upon a receiver is usually a mixture of the signals from multiple independent transmitters. As such, the harvested power at a receiver can be significantly enhanced if the transmit signals can be coordinated
to superimpose each other coherently at the receiver. This is the basic idea of distributed energy beamforming and information transfer (DEBIT), which was  introduced in \cite{Fang15}, where the gain of DEBIT was studied in two-way relay networks. However, such distributed  energy beamforming gain
is quite limited since there are only two transmitters serving
the energy-harvesting relay in a two-way relay network. It is of
pressing interest to understand how this gain scales along
with the number of transmitters, and how it fundamentally
affects the tradeoff between information delivery and power
transfer.

In this work, we apply  DEBIT to  multiway relay networks (MWRNs) \cite{Gunduz13,Amarasuriya12,Yuan14} and
 study the transceiver design to optimize the system performance.    For MWRNs  with a large number of users, a large transmit power budget is usually required at the relay node, especially when AF protocol is adopted by the relay \cite{Fang14,Gunduz13,Amarasuriya12}. It has been shown in \cite{Fang14,Gunduz13} that the throughput  of the  MWRN  over Gaussian channels will be bottlenecked by the relay if it has a limited power supply.  With DEBIT, the relay node can  harvest  power from the  RF signals to improve the network throughput, as well as to prolong the relay's lifetime \cite{Fang14}.

We study the sum-rate maximization and harvested power maximization problems  for the proposed DEBIT scheme. For sum-rate maximization,   the power splitting ratio, the time splitting ratio as well as the power allocations are jointly optimized to maximize the sum-rate of MWRNs under a target harvested power at the relay. The resultant  optimization problem is non-convex, and we show that the global optimal solution can be obtained through a three-dimension search in combination with conventional convex optimization. We also propose a  suboptimal scheme to reduce the complexity, for which the sum-rate maximization problem  becomes   simple linear programming problem. We analyze the   sum-rate performance of the proposed  scheme and a closed-form expression of the  sum-rate lower bound is derived for the MWRN  with a large number of users. In addition, we  also formulate the problem of maximizing the harvested power under the constraint of a target sum-rate for information delivery. We show that this harvested-power maximization problem can be solved by a three-dimensional exhaustive search on top of a convex program. We also propose a low-complexity  suboptimal scheme, and show that   the optimal resource allocation for such a simplified  scheme can be efficiently obtained by  a one-dimensional search  over solutions of linear programs. We further establish an asymptotic lower bound of the average harvested power, and show that the lower bound increases quadratically in the number of users.

The rest of this paper is organized as follows. Section 2 describes the system model and Section 3 describes the DEBIT scheme. Section 4 investigates the sum-rate maximization problem. The achievable sum-rate performance is analyzed in Section 5. Section 6 investigates the harvested power maximization problem. The proposed schemes are tested and compared with the conventional SWIPT scheme in Section 7, followed by the conclusions in Section 8.

\section{System Model}

Fig. 1 (a) shows a half-duplex MWRN  with  $K$ single-antenna users and one single-antenna relay node. We assume there is no direct user-to-user links, and these users need to exchange information via the relay node. Full-data exchange is considered in this work, i.e.,    each user  need to receive the messages from all the other users in two consecutive  phases. Note that such a model corresponds to the case that multiple sensor nodes exchange   information through an intermediate node. In the first phase (referred to as MAC phase), all the   users send signals to the relay. In the second phase (referred to as BC phase), the relay   harvests energy using a power splitter and  then forwards the residual signal to the users using amplify-and-forward relaying protocol. The channel between the $k$-th user and the relay is denoted by  $h_k$,  which is  Rayleigh distributed and keeps unchanged during the two phases. We also assume that each node has perfect knowledge of all the channels.

\section{The Proposed DEBIT Scheme}
\subsection{MAC Phase}
As shown in Fig. 1 (b), the MAC phase  is in general divided into two subphases, and the duration of the two subphases are $\alpha T$ and $(1-\alpha)T$, respectively. Here,   $\alpha \in [0,1]$ and $T$ is the duration of the whole MAC phase.  In the first subphase, as shown in Fig. 1 (b), each user sends superimposed  energy symbols as well as information symbols to the relay. While in the second subphase,   user $k$ transmits information symbols only.

Let $p_{k,E}$ and $p_{k,1}$ denote the transmit power of the  energy  and information symbols at the $k$-th user, respectively. In the first subphase, the transmit signal at user $k$ is given by
\begin{equation}\label{eq.usertx1}
x_{k,1}(t) = \sqrt{ p_{k,E}} e^{-j \angle{h_k}} s_E(t) + \sqrt{p_{k,1}} s_{k,1} (t), t \in [0,\alpha T]
\end{equation}
where  $s_E(t)$ is the energy signal for power transfer,  and $ s_{k,1} (t)$ is the information signal.
%$ \angle h_k $ denotes the phase of $h_k$.

In the second subphase, let  $p_{k,2}$ denote the transmit power of user $k$. Then, the transmit signal of the $k$-th user is given by
\begin{equation}\label{eq.usertx2}
x_{k,2}(t) =  \sqrt{p_{k,2}} s_{k,2} (t),  t \in (\alpha T,T].
\end{equation}

Let $P_k$  denote the transmit power budget at user $k$ in the MAC phase, then $p_{k,E}$, $p_{k,1}$ and $p_{k,2}$ are subject to the following constraint:
\begin{equation}\label{eq.consPU}
\alpha (p_{k,E} + p_{k,1}) + (1-\alpha) p_{k,2} = P_k, \forall k.
\end{equation}

%%%%%%%%%%%%%%%%%%%%%%%%%%%%%%%%%%%%%%%%%%%%%%%%%%%%%%%%%%%%%%
\begin{figure}[t]
\centering
\includegraphics[width=3.7in]{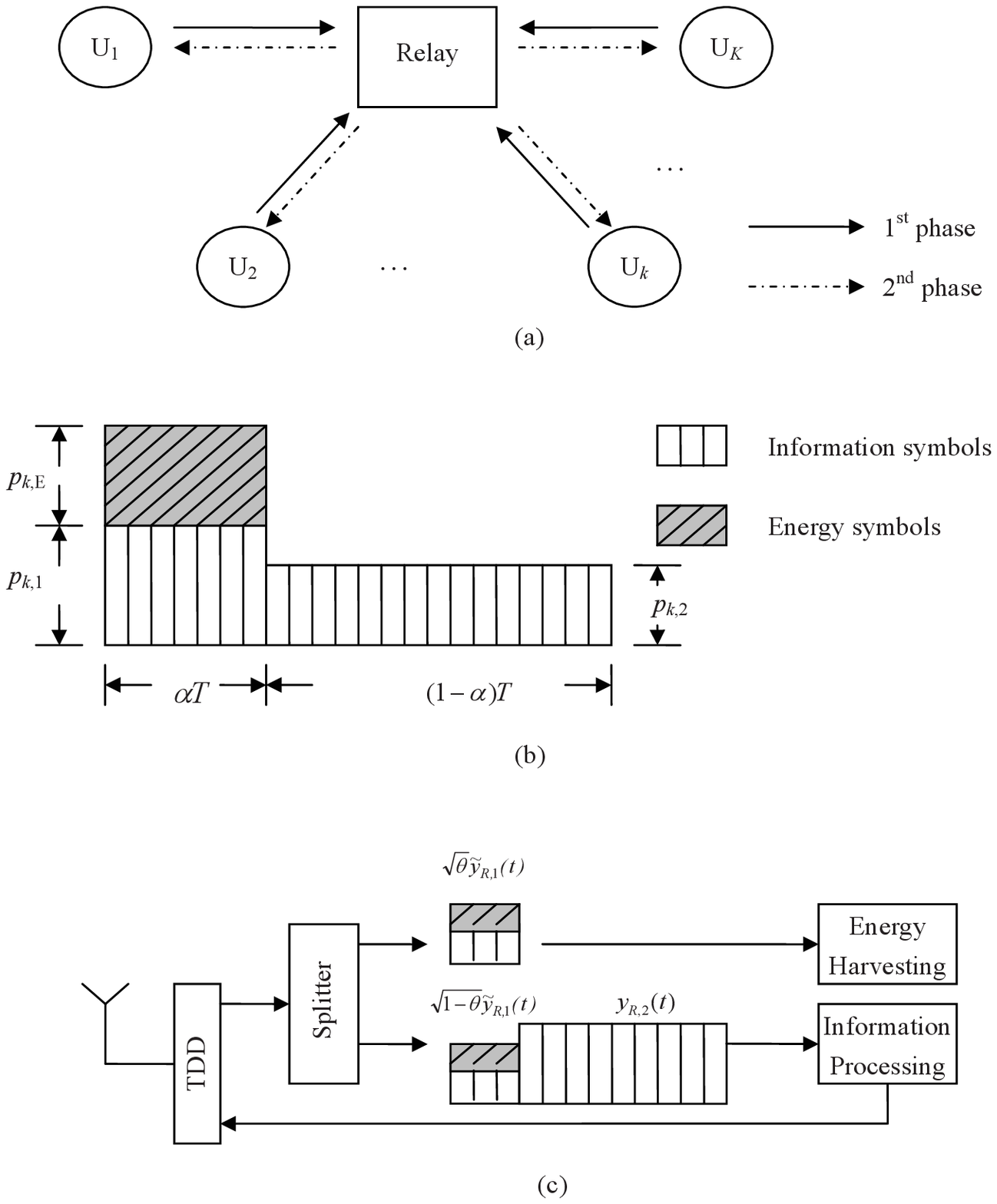}
\caption{(a) A $K$-user multiway relaying system. (b) Frame structure
for the proposed DEBIT scheme. (c) Energy
harvesting and information processing at the relay. }
 \label{fig_sys}
\end{figure}
%%%%%%%%%%%%%%%%%%%%%%%%%%%%%%%%%%%%%%%%%%%%%%%%%%%%%%%%%%%%%%%%%%%%%%%%

\subsection{Relay Energy Harvesting and Information Processing}

From (\ref{eq.usertx1}),  in the first subphase, the received signal  at the relay  is
\begin{equation}\label{relayrx1}
\begin{split}
\tilde{y}_{R,1}(t) &= \sum_{k=1}^K \sqrt{ p_{k,E}}  | h_k| s_E(t) \\
& + \sum_{k=1}^K \sqrt{p_{k,1}} h_k s_{k,1} (t) + z_{R,1,a}(t),
\end{split}
\end{equation}
where $z_{R,1,a}(t) \thicksim {\mathcal CN}(0,\sigma_{R,a}^2)$ is the additive white Gaussian noise (AWGN) at the relay's antenna.

Then, the relay performs  information receiving and energy harvesting based on power splitting \cite{Zhou12}. Specifically,  the received signal $\tilde{y}_{R,1}(t), t \in [0,\alpha T]$, is splitted into two parts: $\sqrt{\theta}  \tilde{y}_{R,1}(t) $ and $\sqrt{1-\theta}  \tilde{y}_{R,1}(t) $, where $\theta \in [0,1]$ is  the relay's power splitting ratio. Here, $\sqrt{\theta}  \tilde{y}_{R,1}(t) $  is used for energy harvesting, and $\sqrt{1-\theta}  \tilde{y}_{R,1}(t) $ is for information forwarding. From (\ref{relayrx1}), the  harvested power in the first subphase is given by
\begin{equation}\label{eq.EH}
\begin{split}
& P_{\text{EH}}(\alpha,\theta,\tilde{\boldsymbol{p}}_E,\tilde{\boldsymbol{p}}_1) =  \alpha  \theta \eta  E(|\tilde{y}_{R,1}(t)|^2) \\
& \quad = \alpha \theta \eta    \left[  \left( \sum_{k=1}^K  \sqrt{p_{k,E}} |h_k| \right)^2 +  \sum_{k=1}^K { p_{k,1}} |h_k|^2 \right],
\end{split}
\end{equation}
where $ \eta \in [0,1]$ is the efficiency of energy conversion , $ \tilde{\boldsymbol{p}}_E  \triangleq[p_{1,E},\ldots,p_{K,E}]^T$, and $ \tilde{\boldsymbol{p}}_1  \triangleq[p_{1,1},\ldots,p_{K,1}]^T$. Note that the first and second term  on the right hand side of (\ref{eq.EH}) denotes the energy harvested from the common  signal $s_E(t)$ and the information signal $\{s_k(t)\}_{k=1}^K$, respectively.

The other part of the received signal, $\sqrt{1-\theta}~ \tilde{y}_{R,1}(t), t \in [0,\alpha T]$, is then converted into baseband for information processing.  It should be noted  that the energy symbols $s_E(t)$ are known to  the relay node. Hence, it can be canceled at the relay before information processing. After  removing the signals related to $s_E(t)$ from $\sqrt{1-\theta}~ \tilde{y}_{R,1}(t)$, the relay obtains
\begin{equation}
\begin{split}
y_{R,1}(t) &= \sqrt{1\!\!-\!\!\theta} \tilde{y}_{R,1}(t)- \sum_{k=1}^K \sqrt{(1\!\!-\!\!\theta) p_{k,E}}  |h_k| s_E  + z_{R,1,b}(t) \\
&= \sum_{k=1}^K \sqrt{(1-\theta) p_{k,1}} h_k s_{k,1} (t) + z_{R,1}(t).
\end{split}
\end{equation}
Here,  $z_{R,1,b}(t) \thicksim {\mathcal CN}(0,\sigma_{R,b}^2)$ is the AWGN introduced by the signal conversion from passband to baseband. $ z_{R,1}(t) \triangleq  \sqrt{1-\theta}z_{R,1,a}(t) + z_{R,1,b}(t)$ with a variance of $\sigma_{R,1}^2 \triangleq (1-\theta) \sigma_{R,a}^2 + \sigma_{R,b}^2$.

As mentioned before, no energy harvesting is involved at the relay in the second subphase, and the  relay's received signal  is given by
\begin{equation}
y_{R,2}(t) =\sum_{k=1}^K \sqrt{p_{k,2}} h_k s_{k,2} (t) + z_{R,2}(t),
\end{equation}
where $t \in (\alpha T, T]$, and  $z_{R,2}(t) \thicksim {\mathcal CN}(0,\sigma_{R}^2)$ is the AWGN.

\subsection{BC Phase}
After energy harvesting and information processing, the relay broadcasts information symbols received in the MAC phase to all the   users in the BC phase. Similarly to the MAC phase, the BC phase consists of two subphases, with the transmit signals  at the relay   expressed as
\begin{equation}
x_{R,n}(t) = \sqrt{\omega}  y_{R,n}(t), ~ n=1,2,
\end{equation}
 where $n$ is a subphase index, and $\omega$ is an amplification factor for power control at  the relay:
\begin{equation}\label{eq.beta}
\begin{split}
 \omega  \times & \left[  \alpha \left((1-\theta) \sum_{k=1}^K  p_{k,1}|h_k|^2  +    \sigma_{R,1}^2 \right) \right. \\
 & \quad \left. +  (1-\alpha) \left( \sum_{k=1}^K p_{k,2}|h_k|^2  +     \sigma_{R}^2 \right)  \right] \le P_R,
 \end{split}
\end{equation}
where $P_R$ denotes the relay power budget.

At user $k$, the received signals in the BC phase can be expressed as
 \begin{equation}\label{eq.ykn}
y_{k,n}(t) = h_k x_{R,n}(t) + z_{k,n}(t),
\end{equation}
where  $z_{k,n}(t) \thicksim {\mathcal CN}(0,\sigma^2)$ is the AWGN.

After removing the self-interference, user $k$ obtains
\begin{equation}\label{eq.yknSI}
\begin{split}
\tilde{y}_{k,n}(t)  &= \sum_{m=1,m \neq k}^K \sqrt{\omega \delta_n p_{m,n}} h_k h_m s_{m,n} (t)  \\
& \quad + \sqrt{\omega}  h_k  z_{R,n}(t) + z_{k,n}(t),
\end{split}
\end{equation}
where $\delta_1 =(1-\theta)$ for the first subphase, and $\delta_2 =1$ for the second subphase.

%\remark The proposed scheme is a generalization of both the conventional power-splitting and time-splitting SWIPT schemes. It specializes to the power-splitting SWIPT with $\alpha=1$, $p_{k,E}=0$,   $p_{k,1}=P_k$, and $p_{k,2}=0, \forall k$, i.e., we only have the first subphase, and all the power is used for information delivery. Also, it specializes to the time-splitting SWIPT with $p_{k,E}=0$, $p_{k,1}=p_{k,2}=P_k, \forall k$, and $\theta=1$, i.e., all the power is used for information delivery and the powers used in the two subphases are the same.

%{\remark Compared with the conventional power-splitting and time-switching SWIPT schemes, a key novelty of the proposed DEBIT scheme is that we  introduce  the energy-harvesting signal $s_E(t)$ in (\ref{eq.usertx1}) for efficient power transfer. Note that $s_E(t)$ is common for every user. Therefore, on one hand, $s_E(t)$  from different users can   coherently superimpose    each other, so as to achieve a distributed energy beamforming gain. On the other hand, the common  energy-harvesting signal $s_E(t)$   cannot carry information, which incurs a certain amount of power inefficiency at the user sides. }

\section{Sum-Rate Maximization}
In this section, we investigate the optimal transceiver design to maximize  the achievable sum-rate of the DEBIT scheme over MWRNs.

\subsection{Optimal Designs}
From (\ref{eq.yknSI}), it can be seen that  the signal model  for the considered    MWRN   represents  a multi-access channel with $K-1$ users. Hence, the achievable rate region for  the proposed DEBIT scheme is the convex hull of
\begin{equation}\label{eq.ARRopt}
\begin{split}
\bigcap_{k=1}^K \bigcap_{{\mathcal S} \in {\mathcal S}_k} &  \left\{ (R_1,\ldots,R_K)  \bigg| \right.  \\
&   \left. \sum_{m \in {\mathcal S}} R_m   \le  R_{k, {\mathcal S}} (\alpha,\theta,\omega,\tilde{\boldsymbol{p}}_1,\tilde{\boldsymbol{p}}_2)  \right\},
\end{split}
\end{equation}
where ${\mathcal S}_k = \{1,...,k-1,k+1,...,K\}$, $\tilde{\boldsymbol{p}}_2 = [p_{1,2},\ldots,p_{K,2}]^T$,   and
 \begin{equation}\label{eq.ARRoptRate}
 \begin{split}
& R_{k, {\mathcal S}}(\alpha,\theta,\omega,\tilde{\boldsymbol{p}}_1,\tilde{\boldsymbol{p}}_2) = \\
& \quad \alpha  {\mathcal C} \left( \frac{ (1-\theta) \omega |h_k|^2\sum_{m \in {\mathcal S}}  |h_m|^2 p_{m,1}}{\omega |h_k|^2((1-\theta) \sigma_{R,a}^2 + \sigma_{R,b}^2 ) + \sigma^2}\right) \\
& \quad + (1-\alpha) {\mathcal C} \left( \frac{ \omega |h_k |^2  \sum_{m \in {\mathcal S}}  |h_m|^2   p_{m,2}}{\omega |h_k|^2 \sigma_{R}^2  + \sigma^2}\right),
\end{split}
\end{equation}
with ${\mathcal C}(x) \triangleq \frac{1}{2} \log_2(1+x)$.

Let $P_{\text{EH}}^0$ denote the target harvested power at the relay. In the following,  we aim to maximize the achievable sum-rate under peak power constraint at each user. The sum-rate maximization problem can be formulated as
\begin{equation}\label{eq.Rsummax}
 \begin{split}
 {\text P1}: \quad &\max_{\alpha,\theta,\omega,\tilde{\boldsymbol{p}}_E,\tilde{\boldsymbol{p}}_1,\tilde{\boldsymbol{p}}_2, \{R_k\}_{k=1}^K} \quad R_{\text{sum}} = \sum_{k=1}^{K}  R_k \\
 &\text{s.t.}~ (\ref{eq.consPU}),(\ref{eq.beta}),(\ref{eq.ARRopt}),  \\
& 0 \le \theta \le 1, \\
& 0 \le \alpha \le 1, \\
 &  P_{\text{EH}}(\alpha,\theta,\tilde{\boldsymbol{p}}_E,\tilde{\boldsymbol{p}}_1)   \ge P_{\text{EH}}^0, \\
 & p_{k,E} + p_{k,1} \le P_{\text{peak}}, k=1,\ldots,K,
 \end{split}
\end{equation}
where $P_{\text{peak}}$  denotes the peak power  at each user.

%Before solving this problem, we first introduce some useful results for the optimal solution of    problem P1.
%
% \lemma   The optimal solution of P1, denoted by $(\alpha^*,\theta^*,\omega^*, \tilde{\boldsymbol{p}}_E^*, \tilde{\boldsymbol{p}}_1^*, \tilde{\boldsymbol{p}}_2^*)$, satisfies
%\begin{equation}\label{eq.alphaopt}
%\frac{P_{\text{EH}}^0}{\eta P_{\text{peak}} \left(\sum_{k=1}^K  |h_k| \right)^2 } \le \alpha^* \le  \frac{\max_k{P_k}}{P_{\text{peak}}},
%\end{equation}
%\begin{equation}
%\frac{P_{\text{EH}}^0} {\eta  \left( \sum_{k=1}^K  \sqrt{P_k} |h_k| \right)^2} \le \theta^* \le 1,
%\end{equation}
%and
%\begin{equation}\label{eq.betaopt}
% \frac{P_R}{\sum_{k=1}^K P_k |h_k|^2 + \tilde{\sigma}_R^2}  \le \omega^* \le \frac{P_R}{\tilde{\sigma}_R^2},
%\end{equation}
%where $\tilde{\sigma}_R^2 = \alpha^* \sigma_{R,1}^2 + (1-\alpha^*) \sigma_R^2$.
%
%\proof See Appendix A.

Note that the function $f(\boldsymbol{\tilde{p}}_E)= \left( \sum_{k=1}^K \sqrt{p_{k,E}} |h_k| \right)^2$ in (\ref{eq.EH}) is concave in $\boldsymbol{\tilde{p}}_E \in {\mathcal R}^{K}$. Furthermore,  the other constraints in (\ref{eq.Rsummax}) are linear constraints when $\alpha, \theta$, and $\omega$ are fixed. Based on these observations, we have the following result.

{\theorem   For  fixed $\alpha, \theta$, and $\omega$, the optimization problem P1 is  convex.}

{\remark Note that both the relay power constraint function (\ref{eq.beta}) and the rate constraints (\ref{eq.ARRoptRate}) are monotonic increasing functions in terms of $\omega$. Hence, with Theorem 1, the global optimal solution for   P1 can be  obtained with a two-dimensional exhaustive search over $\alpha, \theta$, together with a one-dimensional bisection search over $\omega$.}

\subsection{Suboptimal  Designs}
Capitalizing on the distributed energy beamforming gain, the proposed DEBIT scheme can achieve  superior performance over the conventional SWIPT. However, a three-dimensional search is required to find the optimal solution, and it is quite difficult to analyze its performance. In this subsection, we   propose a  suboptimal scheme with low-complexity. We   show that the optimization problem for the  suboptimal scheme  becomes linear-programming problem which are much easier to solve as compared with the optimal scheme.

In this suboptimal   scheme, the first subphase   at each user is dedicated to power transfer, i.e., $p_{k,1}=0, \forall k$, and the   powers used for energy transfer   are equal to each other, i.e., $p_{k,E}=p_E, \forall k$. At the relay node,   the power splitting ratio  is $\theta_{\text{sub}}=1$, as each user transmits energy signals only in the first subphase.    For such suboptimal scheme, the achievable rate region  is the convex hull of
\begin{equation}\label{eq.ARRTS}
\bigcap_{k=1}^K \bigcap_{{\mathcal S} \in {\mathcal S}_k}  \left\{ (R_1,\ldots,R_K) \bigg| \sum_{m \in {\mathcal S}} R_m   \le  R_{k, {\mathcal S},\text{sub}}  \right\},
\end{equation}
where
% \begin{equation}
% \begin{split}
%& R_{k, {\mathcal S},\text{sub}} =  (1-\alpha_{\text{sub}}) \\
%&    \times {\mathcal C} \left( \frac{ \omega_{\text{sub}} |h_k |^2  \sum_{m \in {\mathcal S}}  |h_m|^2   p_{m,2,TS}}{\omega_{\text{sub}} |h_k|^2 \sigma_{R}^2  + \sigma^2}\right).
% \end{split}
%\end{equation}
 \begin{equation}\label{eq.ARRTSRate}
 R_{k, {\mathcal S},\text{sub}} =  (1-\alpha_{\text{sub}})
  {\mathcal C} \left( \frac{ \omega_{\text{sub}} |h_k |^2  \sum_{m \in {\mathcal S}}  |h_m|^2   p_{m,2,TS}}{\omega_{\text{sub}} |h_k|^2 \sigma_{R}^2  + \sigma^2}\right).
\end{equation}

From (\ref{eq.ARRTSRate}), it can be seen that the time duration for energy transfer, $\alpha_{\text{sub}} T$, should be minimized to maximize the achievable data rate in the second subphase. Hence, the power of  the  energy signal in the first subphase should be as high as possible, that is
\begin{equation}
p_{k,E}= p_E = P_{\text{peak}}, ~ \forall k.
\end{equation}

Recall that $\theta_{\text{sub}}=1$ for energy harvesting at the relay node. Then the corresponding harvested power   is given by
\begin{equation}\label{eq.EHTS}
 P_{\text{EH,sub}}(\alpha_{\text{sub}})   = \alpha_{\text{sub}}  \eta    P_{\text{peak}} \left( \sum_{k=1}^K   |h_k| \right)^2.
\end{equation}
For a  target harvested power $P_{\text{EH}}^0$,  we have
\begin{equation}\label{eq.aTS}
\alpha_{\text{sub}} =   \frac{P_{\text{EH}}^0}{\eta   P_{\text{peak}} \left( \sum_{k=1}^K  |h_k| \right)^2 }.
\end{equation}
As a result, the power of $s_{k,2}(t)$ at user $k$ is
\begin{equation}\label{eq.Pk2TS}
p_{k,2,\text{sub}} =  \frac{P_k - \alpha_{\text{sub}} P_{\text{peak}}} {1 - \alpha_{\text{sub}}}, ~ \forall k.
\end{equation}
From (\ref{eq.beta}), (\ref{eq.aTS}) and (\ref{eq.Pk2TS}), the coefficient $\omega$ for the suboptimal DEBIT scheme can be determined by
\begin{equation}\label{eq.betaTS}
 \omega_{\text{sub}} = \frac{P_R }{ (1-\alpha_{\text{sub}}) (\sum_{k=1}^K p_{k,2,\text{sub}} |h_k|^2  + \sigma_{R}^2) }.
\end{equation}

With closed-form expressions of $\alpha_{\text{sub}}$,  $p_{k,2,\text{sub}}$ and  $\omega_{\text{sub}}$ given by (\ref{eq.aTS}),  (\ref{eq.Pk2TS}) and (\ref{eq.betaTS}), respectively, the sum-rate maximization problem for the suboptimal DEBIT scheme can be formulated as
\begin{equation}\label{eq.Rsummax2}
 \begin{split}
{\text P2:}    & \quad \max_{\{R_k\}_{k=1}^K} R_{\text{sum,sub}} = \sum_{m=1}^{K}  R_m \\
 &\text{s.t.}~ \quad \sum_{m \in {\mathcal S}} R_m  \le  R_{k, {\mathcal S},\text{sub}}, \forall S \in S_k, \forall k.
 \end{split}
\end{equation}

From (\ref{eq.Rsummax2}), we see that   P2 is a   linear program, and can be  solved efficiently in polynomial time \cite{Kar84}.

\section{Sum-Rate Performance Analysis}
In this section,   we   focus on analyzing the suboptimal DEBIT scheme, which serves as the performance lower bound of the optimal DEBIT scheme.

\subsection{Sum-Rate Performance Lower Bound}
We consider a symmetric MWRN in which the   power budgets are the same for all the users:  $P_{k} =P$, which implies that $p_{k,2,\text{sub}} = p_{2,\text{sub}}, \forall k$, for the   suboptimal DEBIT. We also assume  $\sigma_R^2 = \sigma^2$, and that the channels are independent and identically distributed (i.i.d.) Rayleigh fading with an average gain of $G$, i.e., $E(|h_k|^2) = G, \forall k$.

Recall that the signal model  in (\ref{eq.yknSI})  represents  a MAC with $K-1$ users, and there are $K$ such multi-access channels in the AF MWRN. Sort the channels in an ascending order as $|h_{\pi(1)}| \le \ldots \le |h_{\pi(K)}|$. For the AF MWRN, the achievable rates of   users $\pi(2),...,\pi(K)$, are bottlenecked  by the worst-channel user $\pi(1)$, while the data rate of user $\pi(1)$ is bottlenecked  by the second worst user $\pi(2)$. Base on these observations, we are able to obtain a sum-rate lower bound for the    DEBIT scheme.

{\theorem   For a symmetric MWRN with a target harvested power of $P_{\text{EH}}^0$, the achievable sum-rate of the proposed suboptimal DEBIT scheme is lower bounded by
 \begin{equation}\label{eq.SumRateLB}
  \begin{split}
 & R_{\text{sum,sub}}^{\text{LB}}(P_{\text{EH}}^0)  =   \\
  & (1-\alpha_{\text{sub}}){\mathcal C} \left( \frac{ \omega_{\text{sub}} |h_{\pi(1)}|^2 \sum_{m=1}^K |h_{m}|^2 p_{2,\text{sub}}} {  \omega_{\text{sub}} |h_{\pi(1)}|^2 \sigma^2   +  \sigma^2  }\right).
 \end{split}
 \end{equation}
}

\proof See Appendix A.

{\remark As a sum-rate lower bound for the suboptimal scheme, the   bound (\ref{eq.SumRateLB}) also serves as the lower bound of the optimal DEBIT  scheme described in the previous Section.}

Based on Theorem 2, we are able to obtain a closed-form   lower bound on the average sum-rate of the DEBIT   scheme.

\proposition For a symmetric   MWRN  with a target harvested power $P_{\text{EH}}^0$, when the number of users are sufficiently large, the average sum-rate of the    DEBIT scheme    is lower bounded by
\begin{equation}\label{eq.RsumAvgTSLB}
\begin{split}
 &   R_{\text{sum}}^{\text{avg,LB}}(P_{\text{EH}}^0)   =  \\
  & \quad {P_{fs}(P_{\text{EH}}^0) (1-\alpha_{\text{sub}}^0)}  {\mathcal C} \left(   \frac{G P_R}{(K + G \omega_{\text{sub}}^0 )(1-\alpha_{\text{sub}}^0) \sigma^2}   \right),
  \end{split}
\end{equation}
where
\begin{subequations}
\begin{align}
& P_{fs}(P_{\text{EH}}^0) = 1- \Phi \left(\frac{c_0}{\sqrt{K}\sigma_H} - \frac{\sqrt{K} \mu_H}{\sigma_H} \right),\\
  & \alpha_{\text{sub}}^0 =  \frac{P_{\text{EH}}^0}{\eta P_{\text{peak}} K  g^2\left(\frac{c_0}{\sqrt{K}},\sqrt{\frac{K \pi G}{4}}, \sqrt{\frac{(4-\pi)G}{4}} \right) },  \\
  & \omega_{\text{sub}}^0 = \frac{P_R }{ (P- a_{TS}^0 P_{\text{peak}}) \sqrt{K} g\left(\frac{c_0^2}{K \sqrt{K}}, \sqrt{K}G,G \right)  },
\end{align}
\end{subequations}
$c_0 =  \sqrt{\frac{P_{\text{EH}}^0}{\eta P}}$, $\mu_H = E[|h_k|] = \sqrt{\frac{\pi G}{4}}$, $\sigma_H^2 = \text{Var}[|h_k|] = \frac{(4-\pi) G}{4}$, $\Phi(x)$ is the cumulative distribution function (CDF) of the  standard  normal distribution ${\mathcal N}(0,1)$, and $g(\zeta,\mu,\sigma) $ is defined as
\begin{equation}\label{eq.gxy}
g(\zeta,\mu,\sigma)   \triangleq    \mu +  \frac{\sigma}{\sqrt{2 \pi} \left[ 1-\Phi \left(\frac{\zeta-\mu} {\sigma} \right) \right]} e^{-\frac{(\zeta-\mu)^2}{2 \sigma^2}}.
\end{equation}

\proof See Appendix B.

\subsection{Discussion}

For the symmetric MWRN with a larger number of users, when the target harvested power satisfies $P_{\text{EH}}^0 \ll \eta \pi K^2PG/4$,   we have $P_{fs}(P_{\text{EH}}^0) \simeq  1$,     $\alpha_{\text{sub}}^0   \simeq 0$, and $\omega_{\text{sub}}^0 \simeq   \frac{P_R}{KPG}$. The lower bound in (\ref{eq.RsumAvgTSLB}) can be further approximated as
\begin{equation}\label{eq.RsumAvgTSLB2}
R_{\text{sum}}^{\text{avg,LB}}(P_{\text{EH}}^0)   \simeq    {\mathcal C} \left(    \frac{KGPP_R}{ (K^2 P + P_R)\sigma^2} \right).
\end{equation}

If the  relay's power budget scales with the number of users, that is, $P_R = KP$, then  from (\ref{eq.RsumAvgTSLB2}) we have
\begin{equation}
\lim_{K \to \infty} R_{\text{sum}}^{\text{avg,LB}}(P_{\text{EH}}^0) =     {\mathcal C} \left(    \frac{PG}{\sigma^2} \right).
\end{equation}
In this case, the   sum-rate lower bound tends to a constant for a large number of users.

On the other hand, if the transmit power of the relay is fixed, then we obtain from (\ref{eq.RsumAvgTSLB2}) that
\begin{equation}
 \lim_{K \to \infty} R_{\text{sum}}^{\text{avg,LB}}(P_{\text{EH}}^0) = 0,
 \end{equation}
that is,   the average   sum-rate  lower bound approaches zero even for a small  $P_{\text{EH}}^0$.

 From the above discussions, we see that   the transmit power of the relay plays an important role on the sum-rate   of  the MWRN. Hence, energy-harvesting at the relay node would be a promising solution to improve the   throughput of  MWRNs with an energy-constrained relay node.

\section{ Harvested Power Maximization}
In this section, we investigate the transceiver design for harvested
power maximization of the  proposed   scheme  under a predetermined
sum-rate constraint.

\subsection{Optimal and Suboptimal Designs}
Let $R_{\text{sum}}^0$ denote the target sum-rate for the MWRN. The harvested
power maximization problem can be formulated as
\begin{equation}\label{eq.PowerMax}
 \begin{split}
 {\text P3}:  ~ &\max_{\alpha,\theta,\omega,\tilde{\boldsymbol{p}}_E,\tilde{\boldsymbol{p}}_1,\tilde{\boldsymbol{p}}_2, \{R_k\}_{k=1}^K} \quad P_{\text{EH}}(\alpha,\theta,\tilde{\boldsymbol{p}}_E,\tilde{\boldsymbol{p}}_1)  \\
 &\text{s.t.}~ (\ref{eq.consPU}),(\ref{eq.beta}),(\ref{eq.ARRopt}),  \\
& 0 \le \theta \le 1, \\
& 0 \le \alpha \le 1, \\
 &   \sum\nolimits_{k=1}^{K}  R_k   \ge R_{\text{sum}}^0, \\
 & p_{k,E} + p_{k,1} \le P_{\text{peak}}, \forall k,
 \end{split}
\end{equation}

Similar to P1, the global optimal solution for P3 can be obtained as follows.

{\theorem   For  fixed $\alpha, \theta$, and $\omega$, the harvested power maximization problem P3 is a convex
problem. That is,  the   optimal solution to  P3 can be   found by  
three-dimensional   searching over the associated convex optimization solutions.}

 To reduce   complexity, we consider the suboptimal scheme as described in Section 4. Then, the harvested-power maximization problem  can be formulated as
\begin{equation}\label{eq.PowerMaxSub}
 \begin{split}
 {\text P4}:  ~ &\max_{\alpha_{\text{sub}}, \{R_k\}_{k=1}^K} \quad {P}_{\text{EH,sub}}(\alpha_{\text{sub}})  \\
 &\text{s.t.}~  0 \le \alpha_{\text{sub}} \le 1, \\
&   \sum_{m \in {\mathcal S}} R_m  \le  \tilde{R}_{k, {\mathcal S}},
\forall S \in S_k, \forall k, \\
 &   \sum\nolimits_{k=1}^{K}  R_k   \ge R_{\text{sum}}^0.
 \end{split}
\end{equation}

For fixed $\alpha_{\text{sub}}$, the above optimization problem   is
a linear program. Hence, the optimal solution of
(\ref{eq.PowerMaxSub}) can be efficiently obtained by   one-dimension
search over $\alpha_{\text{sub}}$.

\subsection{Performance Analysis}
Based on Theorem 2, we are able to obtain an
lower bound on the average  harvested power of DEBIT when there are a large number of users in the MWRN.

{\proposition For the considered symmetric MWRN with  a sufficiently large
number of users,  the average harvested power of
  DEBIT      under a feasible target sum-rate $R_{\text{sum}}^0$
is lower bounded by
\begin{equation}\label{eq.ERavgTS}
{P}_{\text{EH}}^{\text{avg,LB}}(R_{\text{sum}}^0) = \frac{\pi}{4}
\eta G P     \left[  K^2  -   \nu_0  \left(  1 \!\!+\!\!
\nu_0 E_1\left( \frac{\nu_0}{K} \right)  \right) K \right]
e^{-\nu_0},
\end{equation}
where $E_1(x) = \int_{x}^{+\infty} t^{-1} e^{-t}   dt$ is the
exponential integral function, $ \xi_0 =  2^{2 R_{\text{sum}}^0}-1
$, and $ \nu_0  =    {\xi_0 \sigma^2}/{ G P} $.}

\begin{proof}
The proof is similar to that of Proposition 1 and is omitted here.
\end{proof}

{  \remark  Using the inequality that  $E_1(x) \le e^{-x}/x < 1/x$, the lower bound in (\ref{eq.ERavgTS}) can be further bounded by
 \begin{equation}\label{eq.ERavgTS2}
{P}_{\text{EH}}^{\text{avg}}(R_{\text{sum}}^0) \ge \frac{\pi}{4}
\eta   G P   \left[ (1- \nu_0) K^2  -  \nu_0 K \right]
e^{-\nu_0},
\end{equation}
From the above result, we see that the harvested power lower bound increases with the
number of users for a fixed feasible target sum-rate. }

\section{Simulation Results}
In the simulations, the channels are i.i.d., and reciprocal,  and are modeled as $h_k = \sqrt{10^{-2}d^{-3}} \tilde{h}_k$, where $d = 10$  m is the distance between the users and the relay, and $\tilde{h}_k$ is  Rayleigh distributed with unit gain.   Unless otherwise specified, the average transmit power of each user is $P_k=P=30$ dBm, $\forall k$, the relay power is $P_R = K P$,  and   $P_{\text{peak}} = 10P$. The noise variances are set to be $\sigma_{R,a}^2 = \sigma_{R,b}^2= -43$ dBm,  and $\sigma_{R}^2 = \sigma^2 = -40$ dBm, respectively.  The phase duration is normalized as $T=1$ s, and   $\eta = 0.7$ for energy conversion at the relay.

\subsection{Sum-Rate Maximization}
Fig. 2 shows the achievable sum-rate of the proposed DEBIT scheme  with different peak-to-average power ratio (PAPR), which is defined as $\text{PAPR} = 10 \log_{10}(P_{\text{peak}}/P)$. There are $K=8$ users in the MWRN and the target harvested power is $P_{\text{EH}}^0 =-12$ dBm. It is observed that both the optimal and suboptimal DEBIT schemes outperforms the conventional SWIPT scheme significantly. The optimal DEBIT is less sensitive the  PAPR. While for the suboptimal DEBIT scheme, there is certain performance loss for small peak power as compared with the optimal one.

Fig. 3 shows the achievable rate-energy region of a MWRN with $K=8$ users. The   closed-form lower bound (\ref{eq.RsumAvgTSLB}) is also plotted in the figure. From the figure, we see that the gap between the  optimal DEBIT scheme and the suboptimal   scheme is small. It can also be seen that for small   $P_{\text{EH}}^0$, the achievable sum-rate of these schemes are almost the same since only a small fraction of the power is required for energy-harvesting. However, for large harvested power requirement, the   DEBIT schemes outperforms the conventional power splitting SWIPT scheme significantly. For instance, a sum-rate gain of 3 bps/Hz is observed for the optimal DEBIT scheme when $P_{\text{EH}}^0 =-11$ dBm.

%%%%%%%%%%%%%%%%%%%%%%%%%%%%%%%%%%%%%%%%%%%%%%%%%%%%%%%%%%%%%
\begin{figure}[t]
\centering
\includegraphics[width=3.55 in]{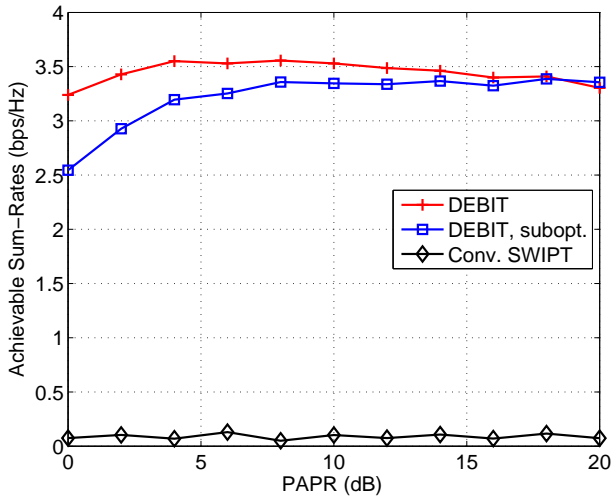}
\caption{ Sum-rate performance   under different peak power limits, $K=8$ users, and $P_{\text{EH}}^0 =-12$ dBm. }
 \label{fig3}
\end{figure}
%%%%%%%%%%%%%%%%%%%%%%%%%%%%%%%%%%%%%%%%%%%%%%%%%%%%%%%%%%%%%%%%%%%%%%%%

%%%%%%%%%%%%%%%%%%%%%%%%%%%%%%%%%%%%%%%%%%%%%%%%%%%%%%%%%%%%%%%%%%%%%%%%
\begin{figure}[t]
\centering
\includegraphics[width=3.55 in]{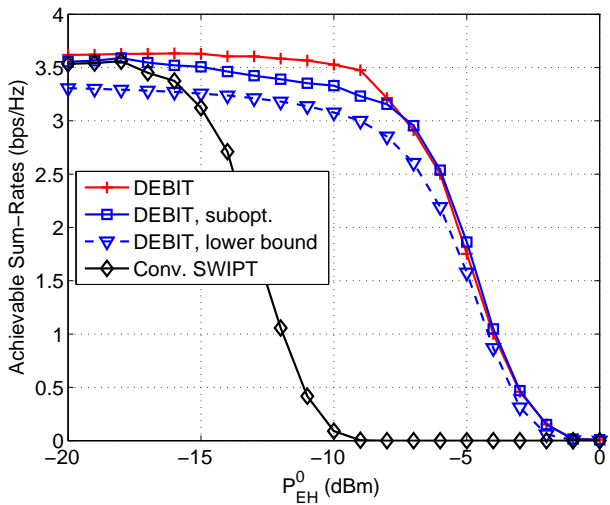}
\caption{ Sum-rate performance   under different target harvested power, $K=8$ users, and $P_{\text{peak}} =10P$. }
 \label{fig2}
\end{figure}
%%%%%%%%%%%%%%%%%%%%%%%%%%%%%%%%%%%%%%%%%%%%%%%%%%%%%%%%%%%%%%%%%%%%%%%%

The sum-rate performance of the proposed DEBIT scheme with different number of users is shown in Fig. 4. The target harvested power is $P_{\text{EH}}^0 = -12$ dBm.   It can be seen that the lower bound becomes tight as the number of user increases. Also,  the  achievable sum-rate  tends to   a constant as the number of users increases, which is consistent with the theoretical analysis.

%%%%%%%%%%%%%%%%%%%%%%%%%%%%%%%%%%%%%%%%%%%%%%%%%%%%%%%%%%%%%
\begin{figure}[t]
\centering
\includegraphics[width=3.5 in]{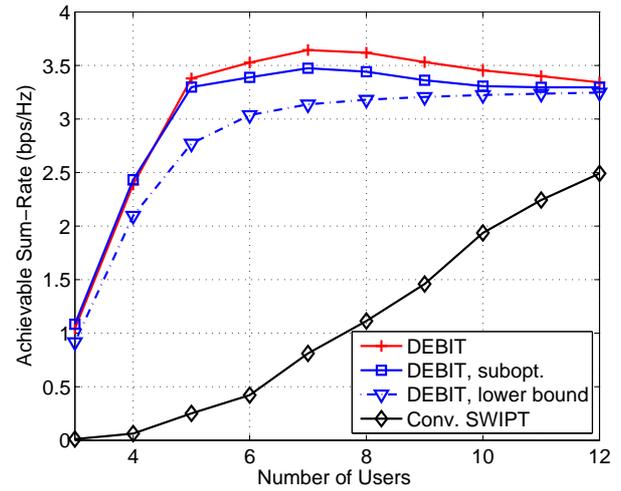}
\caption{ Sum-rate performance   under different number of users,   $P_{\text{EH}}^0 =-12$ dBm, and  $P_{\text{peak}} =10P$. }
 \label{fig4}
\end{figure}
%%%%%%%%%%%%%%%%%%%%%%%%%%%%%%%%%%%%%%%%%%%%%%%%%%%%%%%%%%%%%%%%%%%%%%%%

\subsection{Harvested Power Maximization}

The  average harvested power of the proposed scheme is shown in Fig. 5. The target sum-rate is 2.5 bps/Hz.  From the figure, we see that the derived lower bound becomes very tight for large number of users. It can be also seen that as the number of users increases, more power can be harvested by the relay, and the harvested power   increases much faster than the conventional SWIPT schemes. For instance, in a MWRN with $K=10$ users, the harvested power of  DEBIT   is about eight times as that of the conventional  SWIPT scheme.

%%%%%%%%%%%%%%%%%%%%%%%%%%%%%%%%%%%%%%%%%%%%%%%%%%%%%%%%%%%%%
\begin{figure}[t]
\centering
\includegraphics[width=3.45 in]{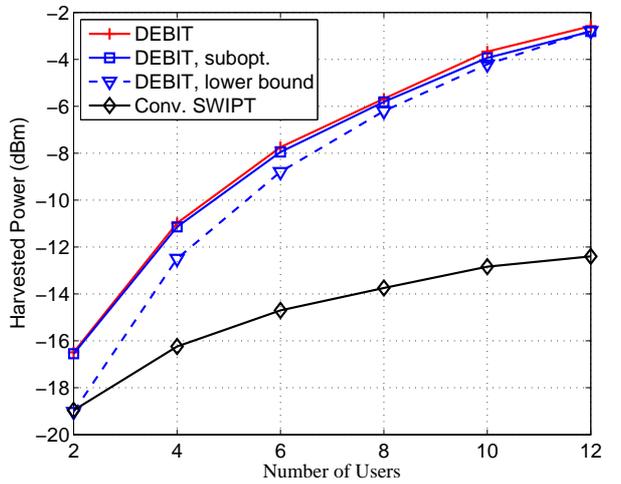}
\caption{Harvested power versus the  number of users,  where the target sum-rate $R_{\text{sum}}^0=$  2.5 bps/Hz. }
 \label{fig3}
\end{figure}
%%%%%%%%%%%%%%%%%%%%%%%%%%%%%%%%%%%%%%%%%%%%%%%%%%%%%%%%%%%%%%%%%%%%%%%%

\section{Conclusion}

This paper proposed a novel DEBIT scheme for efficient energy transfer and information deliver in MWRNs. The sum-rate maximization and harvest power maximization problems were investigated and it was shown that the global optimal solutions for these problems can be obtained with a three-dimensional   search on top of a convex  program.  We also proposed suboptimal schemes to reduce the complexity. In addition,   we analyzed the performance of the  DEBIT scheme and derived closed-form lower bounds of the average sum-rate and the  harvested power.  Numerical results  showed the tightness of the derived lower bounds. The DEBIT scheme establishes a novel joint energy harvesting and information delivery framework, and can be applied to various wireless relay networks or smart-grid powered wireless networks \cite{WangX15,WangX16,Hu16}.

\section{Acknowledgments}
%This work was supported by the China Recruitment Program of Global Young Experts, the National Natural Science Foundation of China (No. 61401400 and No. 61471241), the Natural Science Foundation of Zhejiang Province (No. LY17F010002), and the public welfare project of Zhejiang Province (No. 2016C33036).

This work was supported by   the National Natural Science Foundation of China (No. 61401400 and No. 61471241).

\section{Appendix A: Proof of Theorem 2}
Sort the channels as $|h_{\pi(1)}| \le \ldots \le |h_{\pi(K)}|$. Consider the lower bound first. We assume successive interference cancellation (SIC) is employed at each user with the same decoding order $\pi(1),\pi(2),\ldots,\pi(K)$. Consider decoding at the user $\pi(1)$ with the worst channel gain. From the rate region (\ref{eq.ARRTS}), the achievable sum-rate of the other $K-1$ users is bounded by
\begin{equation}\label{eq.Rworstuser}
\sum_{m=2}^K R_{\pi(m)}  \le R_{\pi(1), {\mathcal S}_{\pi (1)}, \text{sub}}.
\end{equation}
Recall from (\ref{eq.yknSI}) that the channel seen by user $\pi(1)$ (after self-interference cancellation) is a multiple-access channel with $K-1$ users. From the information theory, it is known that the sum-rate in (\ref{eq.Rworstuser}) is indeed achievable using SIC.

For the worst-channel user $\pi(1)$, the achievable rate is bottlenecked  by the second-worst-channel user $\pi(2)$. With the decoding ordered as $\pi(1),\pi(3),\ldots,\pi(K)$ at user $\pi(2)$, the achievable rate of user $\pi(1)$ is
\begin{equation}\label{eq.RworstLB}
\begin{split}
& R_{\pi(1)}^{\text{SIC}} =  (1-\alpha_{\text{sub}})\\
& \times {\mathcal C} \left(\frac{ \omega_{\text{sub}} |h_{\pi(2)}|^2  |h_{\pi(1)}|^2  p_{2,\text{sub}}} {   \omega_{\text{sub}} |h_{\pi(2)}|^2 \left( \sum\limits_{n=3}^K |h_{\pi(n)}|^2 p_{2,\text{sub}} + \sigma^2 \right)  + \sigma^2  }\right).
\end{split}
\end{equation}
From the above two equations,  the achievable  sum-rate  of the proposed suboptimal DEBIT scheme is lower bounded by
\begin{equation}\label{eq.B.RumLB}
 R_{\text{sum,sub}}  \ge R_{\pi(1), {\mathcal S}_{\pi (1)}, \text{sub}} + R_{\pi(1)}^{\text{SIC}}.
\end{equation}
Since $|h_{\pi(2)}| \ge |h_{\pi(1)}|$ and $ |h_{\pi(2)}| \ge 0$, then $R_{\pi(1)}^{\text{SIC}}$ can be lower bounded by
 \begin{equation}\label{eq.RworstLB2}
 \begin{split}
& R_{\pi(1)}^{\text{SIC}}  \ge   (1-\alpha_{\text{sub}})\\
& \times {\mathcal C} \left( \frac{ \omega_{\text{sub}} |h_{\pi(1)}|^4   p_{2,\text{sub}}} {   \omega_{\text{sub}} |h_{\pi(1)}|^2 \left( \sum_{n=2}^K |h_{\pi(n)}|^2 p_{2,\text{sub}} + \sigma^2 \right)  + \sigma^2  }\right).
 \end{split}
 \end{equation}
From  (\ref{eq.B.RumLB}) and (\ref{eq.RworstLB2}), we obtain the sum-rate lower bound in (\ref{eq.SumRateLB}).

\section{Appendix B: Proof of Proposition 1}

From (\ref{eq.consPU}) and (\ref{eq.EH}), the maximum harvested power of the suboptimal DEBIT scheme is
\begin{equation}
P_{\text{EH},{\text{max}}}  =  \eta  P  \left(  \sum_{k=1}^K  |h_k| \right)^2.
\end{equation}
The proposed suboptimal DEBIT scheme is feasible only when $P_{\text{EH},{\text{max}}}    \ge P_{\text{EH}}^0$, that is
\begin{equation}
   \sum_{k=1}^K  |h_k|   \ge c_0 \triangleq  \sqrt{\frac{P_{\text{EH}}^0}{\eta P}}.
\end{equation}
Let $X = \frac{1}{\sqrt{K}} \sum_{k=1}^K  |h_k|$, $\mu_H = E[|h_k|] = \sqrt{\frac{\pi G}{4}}$ and $\sigma_H^2 = \text{Var}[|h_k|] = \frac{(4-\pi) G}{4}$. According to the central limit theorem, $(X - \sqrt{K} \mu_H)$ will converge in distribution to a normal ${\mathcal N}(0,\sigma_H^2)$ for large $K$, i.e.,
\begin{equation}
\lim_{K \to +\infty} \text{Pr} \left[ X - \sqrt{K} \mu_H \le z \right] = \Phi(z/\sigma_H),
\end{equation}
where $\Phi(x)$ is the CDF of the standard normal distribution ${\mathcal N}(0,1)$.  Hence, the feasible probability of the   suboptimal DEBIT scheme can be approximated by
\begin{equation}\label{eq.Pfs}
P_{fs}(P_{\text{EH}}^0) = P\left(X \ge \frac{c_0}{\sqrt{K}} \right) = 1- \Phi \left(\frac{c_0}{\sqrt{K}\sigma_H} - \frac{\sqrt{K} \mu_H}{\sigma_H} \right),
\end{equation}
Next, we consider the achievable sum-rate lower bound when the proposed suboptimal DEBIT scheme is feasible, i.e., $\sum_{k=1}^K  |h_k|   \ge c_0$.  For larger $K$, and under the feasible condition $\sum_{k=1}^K  |h_k|   \ge c_0$, $D_1  \triangleq (\sum_{k=1}^K |h_k|)^2$  can be well approximated by its mean value:
\begin{equation}
 D_1  \simeq  \left[ E\left(\sum\nolimits_{k=1}^K |h_k|   \bigg|   \sum\nolimits_{k=1}^K |h_k|  \ge c_0 \right) \right]^2.
\end{equation}
Since $X = \frac{1}{\sqrt{K}} \sum_{k=1}^K  |h_k|$ is approximately normally distributed for large $K$, $D_1$ can be approximated as $D_1 \simeq K  g^2\left(\frac{c_0}{\sqrt{K}},\sqrt{\frac{K \pi G}{4}}, \sqrt{\frac{(4-\pi)G}{4}} \right)$, where $g(\zeta,\mu,\sigma)$ is defined in (\ref{eq.gxy}).  Similarly, $D_2  \triangleq \sum_{k=1}^K |h_k|^2$  can be well approximated as $D_2 = \sqrt{K} g\left(\frac{c_0^2}{K \sqrt{K}}, \sqrt{K}G,G \right)$ for large $K$.
As a result, for large $K$, the   parameters for the suboptimal DEBIT scheme can be well approximated as
\begin{equation}
a_{\text{sub}} \simeq \alpha_{\text{sub}}^0 =  \frac{P_{\text{EH}}^0}{\eta  P_{\text{peak}} D_1 },
\end{equation}

\begin{equation}
\omega_{\text{sub}} \simeq \omega_{\text{sub}}^0 = \frac{P_R }{ (P- a_{\text{sub}}^0 P_{\text{peak}}) D_2  },
\end{equation}
and
\begin{equation}
p_{2,\text{sub}} \simeq p_{2.TS}^0 =  \frac{P- \alpha_{\text{sub}}^0 P_{\text{peak}}} {1 - \alpha_{\text{sub}}^0}.
\end{equation}
Consequently, the sum-rate lower bound can be approximated as
\begin{equation}\label{eq.C.lb}
R_{\text{sum,sub}}^{\text{LB}}(P_{\text{EH}}^0) \simeq (1-\alpha_{\text{sub}}^0)  {\mathcal C} \left( \frac{ \omega_{\text{sub}} ^0 |h_{\pi(1)}|^2 D_1 p_{2,\text{sub}}^0} {  \omega_{\text{sub}}^0 |h_{\pi(1)}|^2 \sigma^2   +  \sigma^2  }\right).
 \end{equation}
From (\ref{eq.Pfs}) and (\ref{eq.C.lb}), the average sum-rate lower bound for a given target harvested power $P_{\text{EH}} ^{0}$ can be approximated as
\begin{equation}\label{eq.C.RsumApp}
R_{\text{sum,sub}}^{\text{avg,LB}}(P_{\text{EH}}^0)   \simeq  P_{fs}(P_{\text{EH}}^0) E_{\boldsymbol{|h_{\pi(1)}|}}[R_{\text{sum,sub}}^{\text{LB}}(P_{\text{EH}}^0)]
\end{equation}
Note that the PDF of $|h_{\pi(1)}|^2$ is given by $ f_{|h_{\pi(1)}|^2}(x) = \frac{K}{G} e^{-Kx/G}$. From (\ref{eq.C.lb}), (\ref{eq.C.RsumApp}) and after tedious calculation, we can obtain the result in (\ref{eq.RsumAvgTSLB}).


\begin{thebibliography}{99}

\bibitem{DYang15}   Dingcheng, Y.,   Xiaoxiao, Z.,   Lin, X.,   Fahui, W.: 'Energy cooperation in multi-user wireless-powered relay networks', IET Communications, 2015, 9(11), pp. 1412-1420.

\bibitem{Zhai16}  Chao, Z.,  Ju, L.,   Lina, Z.,    Xinhua, W.: 'Wireless energy harvesting-based spectrum leasing with secondary user selection', IET Communications, 2017, 11(4), pp. 499-506.

\bibitem{Varshney08}   Varshney,  L. : 'Transporting information and energy simultaneously',   Proc. IEEE Int. Symp. Inf. Theory (ISIT), Toronto, Canada, July 2008, pp. 1612-1616.

\bibitem{Grover10}  Grover,P.,    Sahai, A. :  'Shannon meets Tesla: Wireless information and power transfer,'   Proc. IEEE ISIT , June 2010, pp. 2363-2367.

\bibitem{Shin16}  Minchul, S.,  Inwhee, J.: 'Energy management algorithm for solar-powered energy harvesting wireless sensor node for Internet of Things', IET Communications, 2016, 10(12), p. 1508-1521.


\bibitem{Mu17}  Yating, W.,  Tao, W.;  Yanzan, S.,  Chongbin, X.: 'Time allocation optimisation for multi-antenna wireless information and power transfer with training and feedback', IET Communications, 2017, 11(3), pp. 414-420.





\bibitem{Zhou12}  Zhou,  X., Zhang, R., and Ho, C. K.: 'Wireless information and power transfer: Architecture design and rate-energy tradeoff', IEEE Trans.   Commun., 2013, 61(11):4754-4767.


\bibitem{Chen16} Xiaojing, C., Wei, N., Xin, W.: 'Optimal Quality-of-Service Scheduling for Energy-Harvesting Powered Wireless Communications ', IEEE Transactions on Wireless Communications, 2016, 15(5):3269-3280.
\bibitem{Chen15} Xiaojing, C., Wei, N., Xin, W. : ' Provisioning quality-of-service to energy harvesting wireless communications ', IEEE Communications Magazine, 2015, 53(4):102-109.
\bibitem{Wang13} Xin, W., Zhaoquan, L. :'Energy-Efficient Transmissions of Bursty Data Packets with Strict Deadlines over Time-Varying Wireless Channels', IEEE Transactions on Wireless Communications, 2013, 12(5):2533-2543.


    \bibitem{Zhang13}  Rui, Z., and  Ho C. K. : 'MIMO Broadcasting for Simultaneous Wireless
Information and Power Transfer,' IEEE Trans. Wireless Commun., 2013, 12(5):  1989-2001.



\bibitem{Wang15} Xin, W., Zheng, N., Tianyi, C. : 'Optimal MIMO Broadcasting for Energy Harvesting Transmitter With non-Ideal Circuit Power Consumption', IEEE Transactions on Wireless Communications, 2015, 14(5):2500-2512.
\bibitem{WangF15} Feng, W., Tao, P., Yongwei, H. : ' Robust Transceiver Optimization for Power-Splitting Based Downlink MISO SWIPT Systems ', IEEE Signal Processing Letters, 2015, 22(9):1492-1496.


\bibitem{Ng13} Ng, D. W. K., Lo, E. S.,  Schober, R.  : 'Energy-efficient resource
allocation in multiuser OFDM systems with wireless information and
power transfer',  IEEE Transactions on Wireless Communications, 2013, 12(12):6352-6370.

\bibitem{Liu16}  Yuanwei, L.,  Lifeng, W.,  Maged, E.,  et al. : 'Two-way relay networks with wireless power transfer: design and performance analysis', IET Communications, 2016, 10(14):1810-1819.


\bibitem{Li13}  Dandan, L., Chao, S.,   Zhengding, Q.: 'Two-way relay beamforming for sum-rate maximization and energy harvesting',   Proc. IEEE ICC, Budapest, Hungary, 2013, pp. 3115-3120.


\bibitem{Tutuncuoglu13} Tutuncuoglu, K.,  Varan, B., and Yener, A. : 'Optimum transmission policies for energy harvesting two-way relay channels',  Proc. IEEE ICC, Budapest, Hungary, June. 2013, pp. 586-590.

\bibitem{Lee13}  Lee, S.,  Rui, Z.,  and Kaibing, H.: 'Opportunistic wireless energy
harvesting in cognitive radio networks',  {\em IEEE Trans. Wireless
Commun.}, vol. 12, no. 9, pp. 4788-4799, Sep. 2013.


\bibitem{Fang15} Zhaoxi, F., Xiaojun, Y., and  Xin, W.: 'Distributed energy beamforming for
simultaneous wireless information and power transfer in the two-way
relay channel', {\em IEEE Signal Process. Lett.}, vol. 22, no. 6, pp. 656-660,
June 2015.

\bibitem{Fang14}  Zhaoxi, F., Xiaojun, Y., and  Xin, W. : 'Distributed energy beamforming and information transfer: A case study for multiway relay channels', Proc. IEEE ICCC 2014, China, 2015:670-675.


\bibitem{Gunduz13} Gunduz, D., Yener, A., Goldsmith, A., and  Poor, H. V. : 'The multi-way relay channel', IEEE Transactions on Information Theory, 2013, 59(1):51-63.

\bibitem{Amarasuriya12} Amarasuriya, G., Tellambura, C., Ardakani, M. :' Performance Analysis of Pairwise Amplify-and-Forward Multi-Way Relay Networks',  IEEE Wireless Communications Letters, 2012, 1(5):524-527.



\bibitem{Yuan14} Xiaojun, Y. : ' MIMO Multiway Relaying With Clustered Full Data Exchange: Signal Space Alignment and Degrees of Freedom,' IEEE Transactions on Wireless Communications, 2014, 13(12):6795-6808.

\bibitem{Kar84} Karmarkar, N.:'  A new polynomial-time algorithm for linear programming,' Combinatorica, 1984, 4:373-395.




\bibitem{Abramowitz70}  Abramowitz, M., and  Stegun, I. : 'Handbook of Mathematical
Functions with Formulas, Graphs, and Mathematical Tables', 9th ed.
National Bureau of Standards, 1970.

\bibitem{WangX15} Xin, W., Yu, Z., Giannakis, G. B., et al. :' Robust Smart-Grid-Powered Cooperative Multipoint Systems,' IEEE Transactions on Wireless Communications, 2015, 14(11):6188-6199.
\bibitem{WangX16} Xin, W., Tianyi, C., Xiaojing, C., et al. :' Dynamic Resource Allocation for Smart-Grid Powered MIMO Downlink Transmissions,' IEEE Journal on Selected Areas in Communications, 2016, 34(12):3354-3365.
\bibitem{Hu16} Shuyan, H., Yu, Z., Xin, W., et al. :' Weighted Sum-Rate Maximization for MIMO Downlink Systems Powered by Renewables,' IEEE Transactions on Wireless Communications, 2016, 15(8):5615-5625.


\end{thebibliography}
\end{document}